
%
\baselineskip 18pt
\magnification=\magstep1
\hsize=15.0truecm
\hoffset=1.0truecm
\vsize=22.0truecm
\font\small=cmr9
\def\spc{\hskip 3 pt}
\hfill {\vbox {\hbox {BARI-TH 178/94} \hbox {\it May 1994}}}
\vskip 1truecm
{\centerline {\bf GENERALIZED GAUSSIAN EFFECTIVE POTENTIAL :}}
\vskip .2 truecm
{\centerline {\bf LOW DIMENSIONAL SCALAR FIELDS}}
\vskip 0.7truecm
{\centerline {Paolo Cea \spc\spc\spc and \spc \spc \spc Luigi Tedesco}}
\vskip 0.5truecm
{\it Dipartimento di Fisica dell'Universit\'a di Bari, I-70126 Bari, Italy
\par
{\centerline {INFN, Sezione di Bari, I-70126 Bari, Italy}}}
\vskip 2.3truecm

{\centerline {\bf ABSTRACT}}

\vskip 1truecm
\par\noindent
We study a generalization of the Gaussian effective potential for
self-interacting scalar fields in one and two spatial dimensions. We compute
the two-loop corrections and discuss the renormalization of the generalized
Gaussian effective potential.

\vfill\eject

The Gaussian effective potential [1]  is usefull to investigate
non-perturbatively
the spontaneous symmetry breaking in scalar field theories. To introduce the
Gaussian effective potential, we recall that the effective potential is the
expectation value of the Hamiltonian in a certain state for which the
expectation value of the scalar field is $\phi_0$ [2]. The Gaussian effective
potential $V_{GEP}(\phi_0)$ is defined by restricting the  states to be trial
Gaussian wave-functionals.
\par
Let us consider a self-interacting real scalar
field $\phi$ in $\nu$ spatial dimensions.

\par\noindent
In the fixed-time Schr\"odinger representation the Hamiltonian is:

    $$H=\int d^{\nu}x \left[-{1\over 2}{\delta^2\over
    {\delta\phi(\vec{x})\delta\phi(\vec{x})}} + {1\over 2}\left(\vec{\nabla}
    \phi(\vec{x}) \right)^2 + {1\over 2} m^2 {\phi^2(\vec{x})} +
    {\lambda \over 4!} {\phi}^4(\vec{x}) \right] \eqno(1)$$
\vskip 0.4truecm
\par\noindent
In the Schr\"odinger representation physical states are wave-functionals.
Let us consider the Gaussian trial wave-functionals centered at $\phi_0$:

$$ \psi_0(\phi)= N \spc exp\left[-{1\over 4}\int d^{\nu}x \spc d^{\nu}y \spc
[ \phi(\vec{x})
- \phi_0 ] \spc  G(\vec{x},\vec{y}) \spc [ \phi(\vec{y}) - \phi_0 ] \spc
\right] \eqno(2)$$

\par\noindent
where

$$ G(\vec{x}, \vec{y}) = \int {d^{\nu}k\over {(2 \pi)^{\nu}}}
e^{i\vec {k} \cdot (\vec {x} - \vec {y})} 2 g(\vec{k}). \eqno(3)$$

\vskip 0.4truecm
\par\noindent
The Gaussian effective potential is defined as:

$$V_{GEP}(\phi_0)={1\over V} min_{|\psi_0>} {<\psi_0| H |\psi_0>\over
{<\psi_0|\psi_0>}}\eqno(4)$$

\par\noindent
where {\it V} is the spatial volume.

\par\noindent
It is a straightforward exercise to show that

$$V_{GEP}(\phi_0)={1\over 2}m^2\phi_0^2 + {\lambda \over {4!}}\phi_0^4 +
{1\over 2}\int {d^{\nu}k\over {(2 \pi)^{\nu}}} g(\vec{k})
- {\lambda\over {32}} \left[
\int {d^{\nu}k\over {(2 \pi)^{\nu}}} {1\over {g(\vec{k})}} \right]^2\eqno(5)$$

\vskip 0.4truecm
\par\noindent
with $g(\vec{k})=\sqrt{{\vec{k}^2} + {\mu}^2}$. The mass $\mu$ satisfies the
gap equation:

$${\mu}^2 = m^2 + {\lambda \over 2}\phi^2_0 + {\lambda \over 4} \int
{d^{\nu}k\over {(2 \pi)^{\nu}}} {1\over {\sqrt{\vec{k}^2 + \mu^2}}}\eqno(6)$$

\vskip 0.4truecm
\par\noindent
Obviously, the Gaussian effective potential goes beyond the perturbation
theory.
However the main disadvantage of the Gaussian effective potential is
the lack of control on the variational extimation Eq.(4). In a previous paper
[3] (henceforth referred to as I), one of us introduced a generalization of the
Gaussian  effective potential which allows to evaluate in a systematic manner
the corrections to the Gaussian approximation. The aim of the present paper
is to evaluate the lowest order corrections to the Gaussian effective
potential.

\par
For reader convenience let us summarize the main results of I.
As a first step,  starting from the ground
state wave-functional $|\psi_0>$,  one sets up a variational basis ${|n>}$.
We stress that the basis is fixed
once and for all by the gap equation (6). This allows us to split the
Hamiltonian as follows

$$H = H_0 + H_I \eqno(7)$$
\vskip 0.3truecm
\par\noindent
where

$$ (H_0)_{nn}=<n|H|n>\eqno(8)$$
$$ (H_I)_{nm}=<n|H|m> \spc\spc\spc\spc  n \spc \ne \spc\spc m \eqno(9)$$

\vskip 0.4truecm
\par\noindent
Note that it is unnecessary to start with a small parameter in $H$ . Indeed the
perturbation Hamiltonian (9) is defined as the off-diagonal elements of
the full Hamiltonian $H$ on the variational basis. In this way we are dealing
with an optimized variational perturbation expansion.

\par\noindent
It turns out that [3]

$$ H_0=E_0(\phi_0,\mu) + \tilde{H_0}\eqno(10)$$

\vskip 0.3truecm
\par\noindent
where

$$E_0(\phi_0,\mu)={<\psi_0|H|\psi_0>\over <\psi_0|\psi_0>},$$
\vskip 0.5 truecm
\par\noindent
and $\tilde {H_0}$ is the normal ordered  Hamiltonian of a free scalar
field with mass $\mu$. Moreover

$$H_I=\int d^{\nu}x \left[ \left(\mu^2 \phi_0 - {1\over 3} \phi^3_0 \right)
:\eta(\vec x): + {\lambda\over {3!}} \phi_0 : \eta^3(\vec x): +
{\lambda\over {4!}} :\eta^4(\vec x): \right], \eqno(11)$$

\vskip 0.5truecm
\par\noindent
with

$$\eta(\vec x)=\phi({\vec x}) - \phi_0. \eqno(12).$$

\vskip 0.4truecm
\par\noindent
After switching on adiabatically the interaction $H_I$, the non-interacting
ground state $|0>$ \spc evolves into the eigenstate of $H$ $|\Omega>$.
Whereupon the natural definition of the
generalized Gaussian effective potential is

$$ V_G(\phi_0)={1\over V}{<\Omega|H|\Omega>\over {<\Omega|\Omega>}}\eqno(13)$$

\vskip 0.4truecm
\par\noindent
with the constraint

$$ {<\Omega|\eta(\vec x)|\Omega>\over {<\Omega|\Omega>}}=0.\eqno(14)$$

\vskip 0.5truecm
\par\noindent
By using the Gell-Mann and Low theorem on the ground state [4], one can show
that [3]:

$$ {V_G(\phi_0)}=V_{GEP}(\phi_0) + {1\over V} \sum_{n=1}^\infty
{(-i)^n\over {n!}}
\int_{-\infty}^0 dt_1...\int_{-\infty}^0 dt_n \cdot$$
$$\cdot <0|T(H_I(0) H_I(t_1)...H_I(t_n)) |0>_{connected}
e^{[\epsilon (t_1+...+t_n)]}.\eqno(15)$$

\vskip 0.4truecm
\par\noindent
In Equation (15) $H_I(t)$ is the interaction Hamiltonian in the
interaction picture.
\par\noindent
A few remarks are now in order.
In the zeroth order in the perturbation, $V_G(\phi_0)$ reduces to the usual
Gaussian effective potential. Higher order corrections to $V_G(\phi_0)$
can be easily evaluated by means of the usual Feynman diagrammatic
expansion.\footnote {\dag}{
\baselineskip 12pt {\small
Note that the constraint (14) sets to zero the tadpole diagrams due to the
linear term in $H_I.$}}
Moreover, the perturbative expansion (15) can be organized into a loop
expansion
likewise the usual perturbative effective potential. The lowest order
correction is given
by the two-loop diagram in Fig. 1. The diagram in Fig. 1 can be easily
evaluated. A straighforward calculation gives

$$ Fig. 1 = -i{{\lambda^2 \phi_0^2}\over 6} \int_{-\infty}^{0}dt\int
d^{\nu}y \spc [\spc i G_F(y)\spc]^3 \spc e^{\epsilon t}, \spc\spc
y=(t,\vec y) \eqno (16)$$

\vskip 0.4truecm
\par\noindent
where

$$G_F(x)=\int {d^{\nu +1}k\over (2 \pi)^{\nu + 1}}
{e^{-ikx}\over {k^2 - \mu^2 +i\epsilon} } \, .\eqno(17)$$

\vskip 0.4truecm
\par\noindent
In the following we shall focus on scalar fields in one and two dimensions.

\par
In one spatial dimension the gap Equation (6) reads

$$x=1+12 \, \hat {\lambda} \, \phi_0^2 - 3 \, {\hat {\lambda}\over {\pi}}\,
lnx \eqno(18)$$

\par\noindent
where $x={\mu^2\over {\mu_0^2}}$, $\mu_0^2=\mu^2(\phi_0=0)$, and
$\hat {\lambda}={\lambda\over {4! \mu_0^2}}$.

\vskip 0.4truecm
\par\noindent
In the lowest order the generalized Gaussian potential is given by the
well-known Gaussian effective potential [1,5]

$${V_{GEP}(\phi_0)\over {\mu_0^2}}=-2 \hat {\lambda} \phi_0^4 +
{{x-1}\over {24 \lambda}}\left[1+{{3 \hat {\lambda}}\over \pi} + {{x-1}\over 2}
\right] \, .\eqno(19)$$

\vskip 0.4truecm
\par\noindent
In Eq.(19) we have subtracted $V_G(\phi_0)$. Thus there is spontaneous
symmetry breaking if $V_G(\phi_0)$ develops a negative minimum away from
the origin. We would like to stress that the ultraviolet divergences
have been absorbed into $\mu_0$:

$$\mu_0^2=m^2 + {\lambda\over 2} \int_{0}^{\Lambda}dk
{1\over {\sqrt{\vec {k}^2 + \mu_0^2}}}\eqno(20)$$

\vskip 0.4truecm
\par\noindent
where $\Lambda$ is an ultraviolet cut-off. Indeed, in one space and one time
dimension we only need to renormalize the mass. In other words, a
suitable definition of the bare mass makes finite the phisical mass
defined as (minus) the irreducible 2-point function at zero momentum.

\par\noindent
Equations (19) and (20) show that the generalized Gaussian effective potential
is free from ultraviolet divergences after mass renormalization on the
$\phi_0=0$ vacuum. Indeed  in the approximation of Eq. (19) the $\phi_0=0$
vacuum is the free vacuum of the scalar field with mass $\mu_0$.
So we have in this approximation
$$ m_{phys}=\mu_0 \, .\eqno(21)$$

\vskip 0.4truecm
\par\noindent
In fig. 2 we  show $V_{GEP}(\phi_0)$ as a function of $\phi_0$ for
various values of $\hat {\lambda}$. Several features are worth mentioning.
Firstly, $\phi_0=0$ is always a local minimum of $V_{GEP}(\phi_0) $.
For $\hat {\lambda}<\hat {\lambda}_c$, $\hat {\lambda}_c \simeq 2.5527$,
the $\phi_0=0$ vacuum is the true ground state. On the other hand, for
$\hat {\lambda}>\hat {\lambda}_c$ the ground state is for $\phi_0\ne0$.

\par\noindent
Thus, a first order phase transition occurs at $\hat {\lambda}_c$. However,
S. J. Chang [6] pointed out that the Simon-Griffiths theorem [7] rules
out the possibility of a first order phase transition in the one
dimensional $\lambda \phi^4$ field theory. In addition, Chang showed
[6] that there is no contradiction between the existence
of a second-order transition and the Simon-Griffiths theorem.
Remarkably it turns out that the lowest order correction Eq. (16) gives
rise to a second-order phase transition. To see this, we note that the
correction (16) is finite:

$$Eq. (16)=-a \, {{\hat {\lambda}^2}\over x} \, \phi_0^2 \, \mu_0^2,
\, \, \, a\simeq 0.7136\eqno(22)$$

\vskip 0.4truecm
\par\noindent
so that

$${V_G(\phi_0)\over {\mu_0^2}} = {V_{GEP}(\phi_0)\over \mu_0^2} -
{\hat {\lambda^2} \phi_0^2\over x}\,  a \, . \eqno(23)$$

\vskip 0.4truecm
\par\noindent
In Fig. 3 we display Eq. (23). We see that there is a second-order
phase transition at
$\hat {\lambda}_c \simeq 0.8371.$ This is confirmed by considering the
mass-gap of the $\phi_0=0$ vacuum. One can easily check that

$$m^2_{phys}=\left.{{\partial^2 V_G(\phi_0)}\over
{\partial \phi_0^2}}\right|_{\phi_0=0} \, .\eqno(24)$$

\vskip 0.4truecm
\par\noindent
We would like to stress that equation (24) is not trivial, because $m^2_{phys}$
is defined by means of the zero-momentum 2-point function. A remarkable
consequence of Eq.(24) is that the mass renormalization of the Gaussian
effective potential extends to $V_G(\phi_0)$.

\par\noindent
{}From Equations (23) and (24) it follows

$${m^2_{phys}(\hat {\lambda})\over {\mu_0^2}}=\left(1-{\hat {\lambda}\over
{\hat {\lambda}_c}}\right), \spc \spc \hat {\lambda} \le \hat
{\lambda}_c \, .\eqno(25)$$

\vskip 0.4truecm
\par\noindent
Our results are in agreement with previous studies [8,9]. However our
generalized Gaussian effective potential relies on a firm field theoretical
basis. In particular Eq.(15) allows us to take care of the higher order
corrections. As a matter of fact we have checked that the higher order
corrections to Eq. (23) do not modify the order of the transition. A full
account of this analysis will be presented elsewhere.

\par\noindent
Moreover in our scheme there are not ambiguities in the renormalization of
ultraviolet divergencies. To illustrate this last point let us consider the
scalar fields in two spatial dimensions. Here, at variance of
the previous case,
the two-loop correction is
logarithmically divergent. The gap equation becomes:

\vskip 0.4truecm
\par\noindent
$$\sqrt x = -3{\hat {\lambda}\over {2 \pi}} +
\sqrt {\left(1+{{3 \lambda}\over {2 \pi}}\right)^2 + 12 \hat {\lambda}
\Phi_0^2}\eqno(26)$$

\vskip 0.4truecm
\par\noindent
where, now, $ \, x={\mu^2\over {\mu^2_0}} \; , \;
 \hat {\lambda}={\lambda\over {4! \mu_0}} \,$ , and
 $\Phi_0^2={\phi_0^2\over\mu_0} \, $.

\par\noindent
The Gaussian effective potential is easily obtained [1]

$${V_{GEP}(\Phi_0)\over {\mu_0^3}}={\Phi_0^2\over2} + \hat {\lambda}\Phi_0^4 -
{(\sqrt x - 1)^2\over {24 \pi}}\left[1+{9\over {2 \pi}}\hat {\lambda}
+2 \sqrt x \right].\eqno(27)$$

\vskip 0.4truecm
\par\noindent
Also in two spatial dimensions the gaussian effective potential
 displayes a first order phase transition at $\hat {\lambda}=\hat {\lambda}_c
\simeq 3.0784$.

\par\noindent
As concern the two-loop contribution Eq.(16), we get

$$Eq.(16)=\mu_0^3 \, {3\over \pi^2} \,
{\hat {\lambda}}^2 \, \Phi_0^2 \, [\,ln(\mu \epsilon) + \gamma + ln3 +
O(\epsilon^2)\,]\eqno(28)$$

\vskip 0.4truecm
\par\noindent
where  $\epsilon$ is the the short distances cut-off  and $\gamma$ is the
Euler's constant.

\par\noindent
The logarithmic divergence in (28) is eliminated by renormalizing the mass.
We have

$$m^2_{phys}=-\Gamma^{(2)}(0)=\mu_0^2+{{\lambda^2 }\over
{192 \pi^2}}[ \, ln(\mu_0 \epsilon) + \gamma + ln3 + O(\epsilon^2) \, ]
 \eqno(29)$$

\vskip 0.4truecm
\par\noindent
By using Eq. (29) and the gap equation for $\phi_0=0$, one can eliminate the
bare mass in favour of the physical mass.
Inserting into $V_G(\phi_0)$ one gets the finite result:

$${{V_G(\Phi_0)\over {\mu_0^3}}}={1\over 2}\Phi_0^2 + \hat {\lambda} \Phi_0^4 -
{{(\sqrt x - 1)^2}\over {24 \pi}}\left[1+{9\over {2 \pi}} \hat {\lambda} +
2\sqrt x \right] +
{3\over {\pi^2}} \, {\hat {\lambda}}^2 \, \Phi_0^2 \, ln {\sqrt x} \eqno(30)$$

\vskip 0.4truecm
\par\noindent
In equation (30) we assumed ${m^2_{phys}\over \mu_0^2}=1$.

\par\noindent
In figure 4 we display $V_G(\Phi_0)$. As it is evident there is no spontaneous
symmetry breaking. Our result contrasts with Ref.[10]. Indeed in Ref. [10] it
is
suggested  that a second order phase transition develops. The discrepancy
resides in the different use of the gap equation.
In our scheme the gap equation is fixed once and for all. On the other hand,
in Ref.[10] the gap equation is fixed at each order by the principle of minimal
sensivity [11]. In this way, however,  there are ambiguities in the
renormalization of the effective potential.
\par\noindent
The main advantages of our generalized Gaussian effective
potential are implicit in the definition Eqs. (13) and (14).
Indeed  these equations put our generalized Gaussian effective potential on the
same level of the perturbative effective potential. Moreover, Eq.(15) allows
a diagrammatic expansion of the higher order corrections which is amenable
to a diagrammatic resummation.

\par\noindent
Finally, as we have already discussed, the renormalization of
the generalized Gaussian
effective potential relies on the underlying $\phi_0=0$ field theory. As a
consequence there are no ambiguities in the renormalization. We hope
to return on this last point for the more interesting case of scalar fields
in three spatial dimensions.

\vfill\eject

{\bf FIGURE CAPTIONS}
\vskip 1truecm
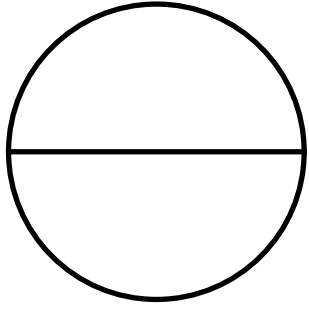
\item {{\bf Fig. 1}} Lowest order correction to the Gaussian effective
potential.
\vskip 1truecm
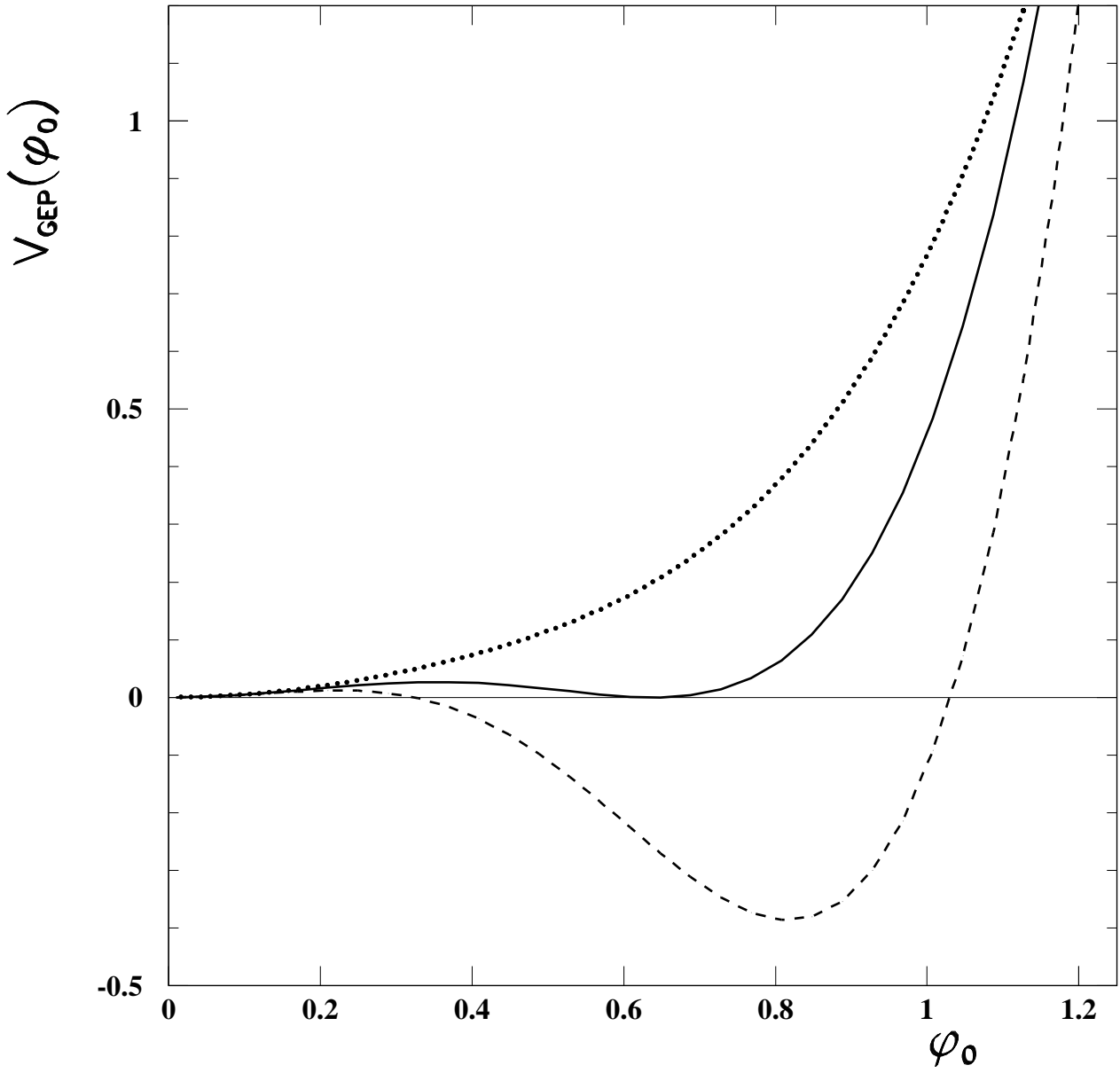
\item {{\bf Fig. 2}} The Gaussian effective potential for $\nu=1$:
 $\hat {\lambda}=1.0$ (dotted line), $\hat {\lambda}=\hat {\lambda}_c$
(solide line) and $\hat {\lambda}=4.0$ (dashed line).
\vskip 1truecm
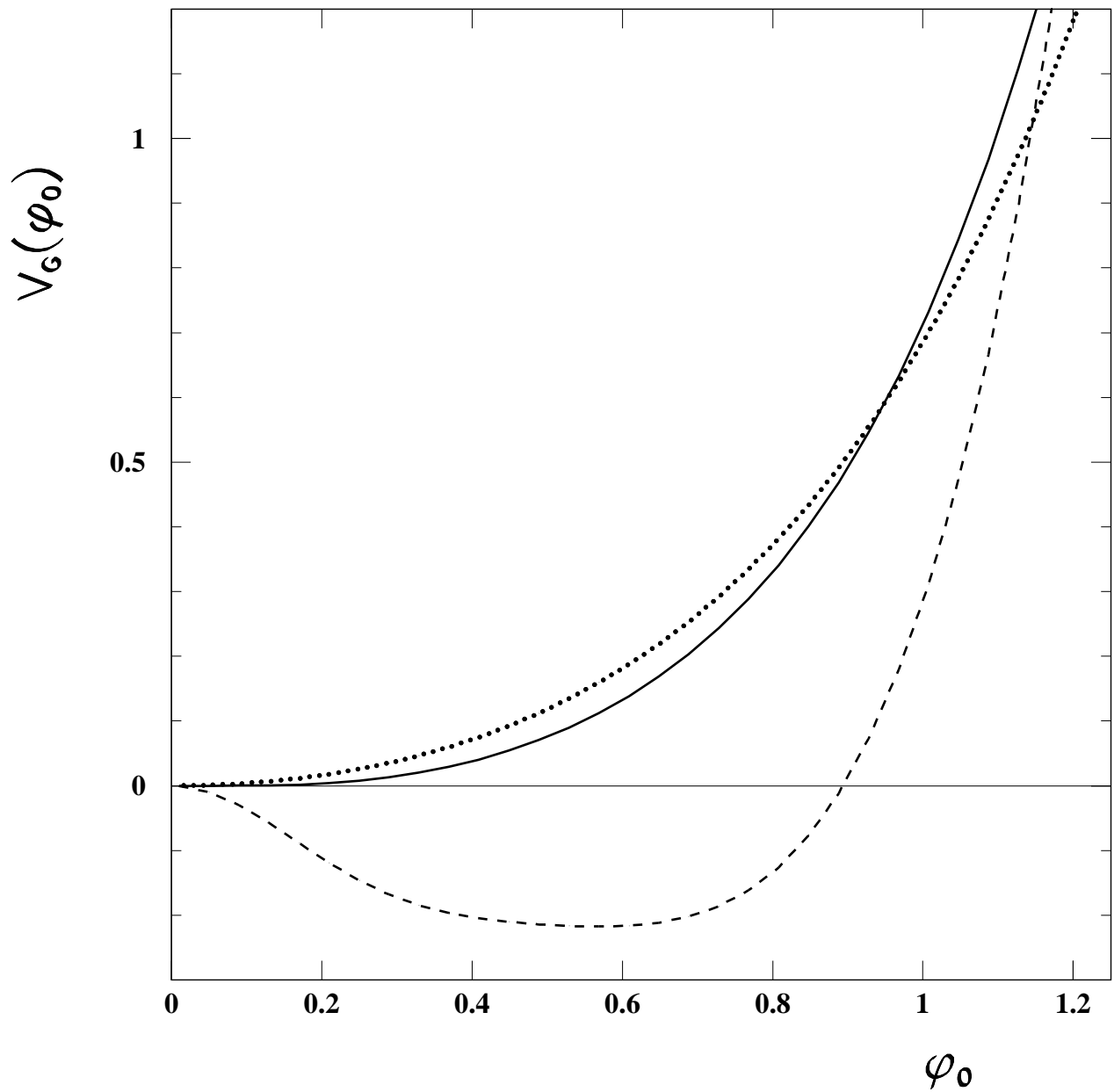
\item {{\bf Fig. 3}} The two-loop generalized Gaussian effective potential for
$\nu=1$.
\item {} Dotted line refers to $\hat {\lambda}=0.4$, solide line to $\hat
{\lambda}=\hat {\lambda}_c$, and dashed line to $\hat {\lambda}=2.5$.
\vskip 1truecm
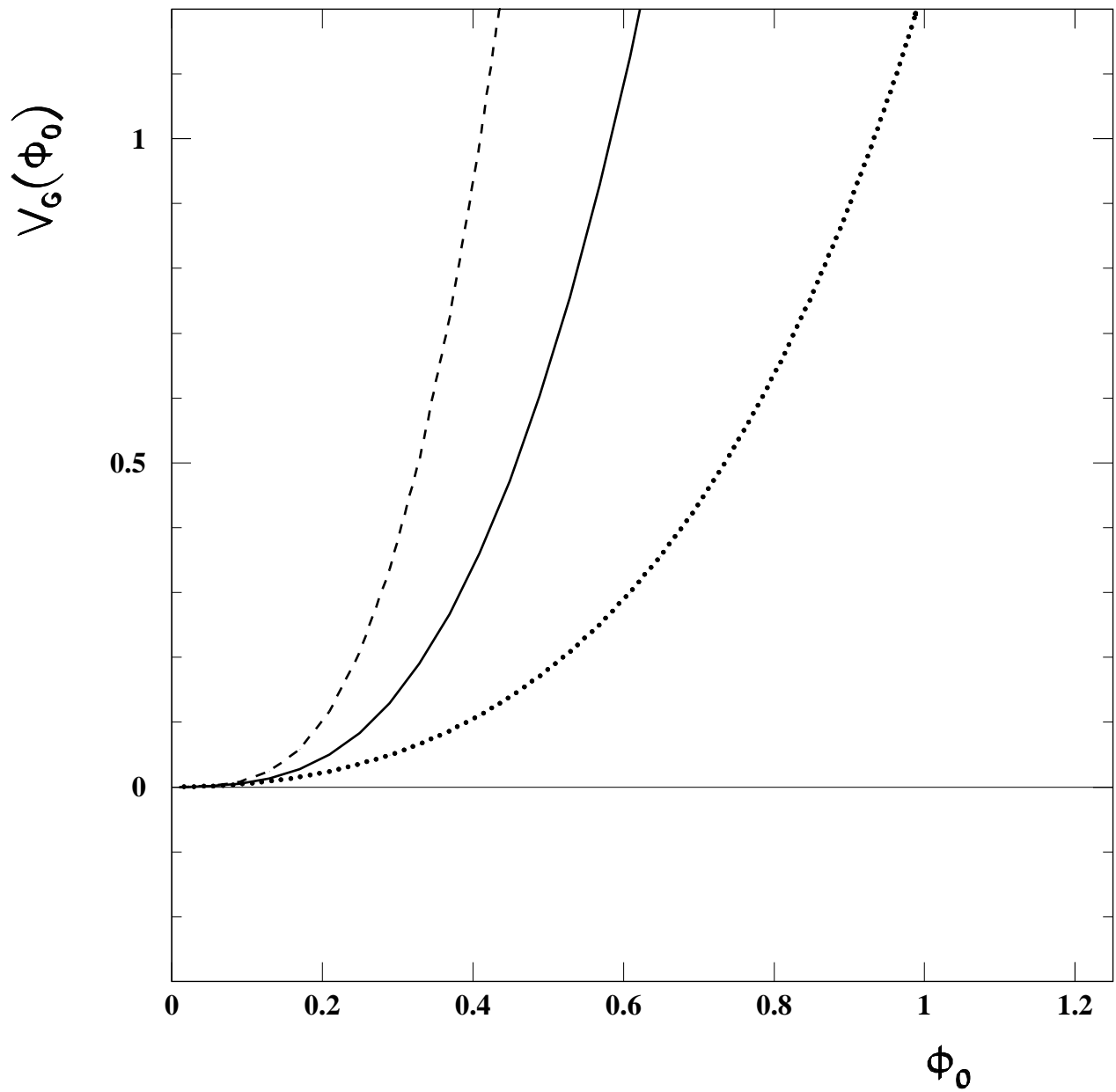
\item {{\bf Fig. 4}} The two-loop generalized Gaussian effective potential
in two spatial dimensions for
$\hat {\lambda}=1$ (dotted line), $\hat {\lambda}=3.0$ (solide line),
and $\hat {\lambda}=5.0$ (dashed line).

\vfill\eject

{\centerline {\bf REFERENCES}}
\vskip 1. truecm
\item {[1]} P.M. Stevenson, Phys Rev. {\bf D30} (1984) 1712; {D32} (1985) 1389;
{\bf D33} (1986) 2305, and references therein.
\vskip 0.3truecm
\item {[2]} G. Jona-Lasinio, Nuovo cimento {\bf 34} (1964) 1790;
\item {}    K. Symanzik, Com. Math. Phys. {\bf 16} (1970) 48;
\item {}    S. Coleman, in Law of Hadronic Matter, ed A. Zichichi (Academic
Press, New York, 1975);
\vskip 0.3truecm
\item {[3]} P. Cea, Phys Lett. {\bf B236} (1990) 191;
\vskip 0.3truecm
\item {[4]} M. Gell-Mann and F. Low, Phys. Rev. {\bf 84} (1951) 350;
\item {}    see also: A.L. Fetter and Walecka, Quantum Theory of
            Many-Particle Systems (Mc. Graw-Hill, New York, 1971;
\vskip 0.3truecm
\item {5]} S. J. Chang, Phys. Rev. {\bf D12} (1975) 1071;
\vskip 0.3truecm
\item {[6]} S.J. Chang, Phys. Rev. {\bf D13} (1976) 2778;
\vskip 0.3truecm
\item {[7]} B. Simon and R.B.Griffiths, Commun. Math. Phys. {\bf 33} (1973)
145;
\vskip 0.3truecm
\item {[8]} L. Polley and U. Ritschel, Phys Lett. {\bf B221} (1989) 2778;
\vskip 0.3truecm
\item {[9]} M. H. Thoma, Zeit. Phys. {\bf C44} (1989) 343;
\vskip 0.3truecm
\item {[10]} I. Stancu, Phy. Rev. {\bf D43} (1991) 1283;
\vskip 0.3truecm
\item {[11]} I. Stancu and P.M. Stevenson, Phys Rev. {\bf D42} (1990) 2710.

\bye